\documentclass[conference]{IEEEtran}

\usepackage{latexsym}
\usepackage{booktabs}

\usepackage{mathtools}
\usepackage{amsmath,amssymb,bm}

\usepackage[footnotesize]{subfigure}
\usepackage{epic}
\usepackage{eepic}
\usepackage{enumerate}

\usepackage{amsthm}
\usepackage{vinayak}
\usepackage{graphicx, color}
\usepackage{verbatim}
\usepackage[mathscr]{eucal}
\usepackage{xspace}
\usepackage{stmaryrd}
\usepackage{paralist}
\usepackage{enumitem}

\usepackage{tikz}
\usetikzlibrary{arrows,backgrounds,decorations,decorations.pathmorphing,positioning,fit,automata,shapes,snakes,patterns,plotmarks,calc}

\makeatletter
\newif\if@restonecol
\makeatother

\numberwithin{equation}{section}
\begin{document}

\newcommand{\RM}[1]{{\textcolor{red}{ \textbf{RM:} #1 }}}
\newcommand{\VP}[1]{{\textcolor{blue}{ \textbf{VP:} #1 }}}
\newcommand{\JD}[1]{{\textcolor{green}{ \textbf{JD:} #1 }}}




\title{
Parameter Optimization in  Control Software using Statistical Fault Localization  Techniques
}
\author{
\IEEEauthorblockN{Jyotirmoy Deshmukh} 
\IEEEauthorblockA{University of Southern California\\
\texttt{jyotirmoy.deshmukh@usc.edu}
}
\and
\IEEEauthorblockN{Xiaoqing Jin}
\IEEEauthorblockA{
\texttt{jinx@cs.ucr.edu}
}
\and
\IEEEauthorblockN{Rupak Majumdar}
\IEEEauthorblockA{MPI-SWS\\
\texttt{rupak@mpi-sws.org}
}
\and
\IEEEauthorblockN{Vinayak S. Prabhu}
\IEEEauthorblockA{MPI-SWS\\
\texttt{vinayak@mpi-sws.org}
}
}

\maketitle

\begin{abstract}
Embedded controllers for cyber-physical systems are often
parameterized by look-up maps representing discretizations of
continuous functions on metric spaces.  For example, a non-linear
control action may be represented as a table of pre-computed values,
and the output action of the controller for a given input computed
by using interpolation. For industrial-scale control systems, several
man-hours of effort are spent in tuning the values within the look-up
maps.
Suppose that during testing,
the controller code is found to have sub-optimal performance.  The
parameter fault localization problem asks which parameter values in
the code are potential causes of the sub-optimal behavior.  We present
a statistical parameter fault localization approach based on
\emph{binary similarity coefficients} and \emph{set spectra} methods.
Our approach extends previous work on (traditional) software fault localization to a
quantitative setting where the parameters encode continuous functions
over a metric space and the program is reactive. 

We have implemented our approach in a simulation workflow for
control systems in Simulink.  Given controller code with
parameters (including \emph{look-up maps}), our framework bootstraps
the simulation workflow to return a ranked list of map entries which
are deemed to have most impact on the performance.  On a suite of
industrial case studies with seeded errors, our tool was able to
precisely identify the location of the errors.

\end{abstract}

\section{Introduction}

The correct operation of modern cyber-physical systems relies on a
complex software controller interacting with a physical plant (\ie, a
model of the physical processes that we wish to control).  Controllers
used in industrial-scale software present a formidable challenge for
formal analysis techniques due to their scale and format. Typical
controllers are defined over several modes of operation, each of which
may use a customized control scheme. Control schemes based on
feedforward maps that seek to statically cancel non-linearities are
frequently used \cite{bohn2006optimization,chengao}. Even when
feedback control schemes such as PID control are employed, it is
common for the proportional, integral and derivative gains to vary
across different modes of operation.  As a result, embedded control
software typically has a fixed high-level structure corresponding to
the ``control law'' chosen by the designer. Flexibility in the
algorithm is offered instead by lookup-maps representing nonlinear
functions or mode-specific control rules. A typical design process
then involves careful hand-tuning of these maps by engineers till the
overall system meets the desired performance objectives.  
As a result, most industrial-scale controllers do not have closed-form
symbolic representations, but have parametric representations such as
 explicit \emph{look-up maps}.  A look-up map $\lutmap$ is a
function from a finite subset of $\reals^d$ to $\reals^n$ (\eg, a
look-up map in two dimensions is a two dimensional table with table entries 
in $\reals^n$). It defines
a mathematical function $f_\lutmap^{\interpol}: C \rightarrow
\reals^n$, for $C$ a bounded subset of $\reals^d$, by completing the
map $\lutmap$ according to the specified interpolation scheme
$\interpol$ (\eg, linear, bilinear, bicubic interpolation etc.).
Whenever controller performs a ``look-up'' for the value $\entry$,
\ie, wishes to compute $f_\lutmap^{\interpol}(\entry)$, the control
software computes the value using the stored map $\lutmap$ under the
interpolation scheme $\interpol$.

Suppose that we are given the code for a controller, with one or more
look-up maps, and on simulating the controller in closed-loop with the
plant, we find that there is a violation of some correctness or
performance specification.  Which map entries should we consider for
investigation?  Which map entries involve performance critical regions
of the controller?  The problem of identifying the entries most
indicative of error or performance loss is called \emph{ fault
localization}, and tools for effective fault localization are key to a
fast design debugging and optimization process.

%


In this paper, we derive a (heuristics based) ranking on the estimated
importance of map entries to an observed failure or to overall system
performance.  If map entry $\entry$ is ranked higher than $\entry'$,
then performance can likely be improved by changing  $\entry$ rather
than $\entry'$.  Our approach obtains this ranking by computing
\emph{binary similarity coefficients} and using \emph{set spectra}
based methods.  A binary similarity coefficient $\phi(X,Y)$  assigns a
measure of similarity between two binary categories $X$ and $Y$ for
pattern classification~\cite{Choi10,Jackson89}.  In our setting, $X$
is the category of failed or unfavorably performing  executions, and
$Y$ is the category of executions that ``look-up'' an entry $\entry$
or its neighborhood.  The binary similarity based ranking approach
computes $\phi(X,Y)$ for each entry $\entry$ in the map, and orders
entries based on the $\phi(X,Y)$ values.  This approach has the
advantage that no complicated controller analysis is performed.  It
integrates with existing simulation-guided analysis tools, and we only
look at  the similarity between failed runs and map entry access in a
black box fashion.  Set spectra based methods define sets of interest
using set algebra methods, also in a black box fashion.

Note that the black-box assumption for industrial control systems is
crucial: plant models for real-world physical systems are routinely
modeled as hybrid and nonlinear differential or algebraic equations,
and controllers have considerable complexity. Traditional verification
tools such as model checkers or reachability analysis tools are unable
to digest the scale and complexity of such models.

Our approach is inspired by earlier work on software fault
localization~\cite{Jones05,LiblitNZAJ05,Liu2005,Jiang2007,Jeffrey08,Naish2011,Robetaler12,Yoo2013}
The similarity coefficient based approach is motivated by the
work~\cite{Jones05,WongDGL14} in software localization, and the set
spectra based methods by the
work~\cite{Pan92heuristicsfor,RenierisR03}.  We extend and improve the
ideas in these works in two ways to suit our context.  First, we
observe that compared to normal software where runs are only
classified as being good or bad, executions for cyber-physical systems
are ascribed numerical scores (e.g., performance measures or
robustness w.r.t.\ logical specifications~\cite{DonzeM10,DonzeFM13}).
Second, in most instances, a look-up map $\lutmap$ represents a
discrete approximation of the associated \emph{continuous} function
$f_\lutmap^{\interpol}$.  Consequently, in most cases, the map value
$\lutmap(\entry)$ is close to the map value $\lutmap(\entry')$ if
$\entry$ and $\entry'$ are close.  This implies that an execution can
indirectly contribute to the importance level of a map entry $\entry$
even if that execution does not directly access $\lutmap(\entry)$ --
this happens when indices close to $\entry$  are accessed.  This
continuity property is absent in traditional software where
individual statements that are problematic can be spread out far apart
from each other.  We show how to cleanly extend the two Boolean
software fault localization approaches to account for the additional
structure in our domain by defining the rankings using certain basic
quantitative functions which give a measure of the effect of a map
entry on a given run.

The problem of fault localization for look-up maps 
can be viewed as a parameter optimization or
tuning problem, where we are trying to determine values for parameters (in this case
the map entries) such that the value of a certain cost function (in this
case the quantitative score indicating well-behavedness of the output) is maximized.
In this sense, black-box optimization approaches can be theoretically used to
perform fault localization. However, even a relatively small 20x20 
look-up map  contains 400
entries, or from the perspective of the black-box optimizer, 
a large 400-dimensional space
over which it needs to optimize. Such an approach, {\em if it scales}, would be
appropriate to synthesize a new look-up map in entirety to make the design
satisfy its specification, but it is not clear whether it is well-suited to
determining which entries in an {\em existing} look-up map are responsible for
making the design not satisfy its specification (identifying sub-performing regions
is of key importance to engineers). Other parameter tuning tools such
as the PID tuning capability within the Control System toolbox in Simulink have very
specific tuning abilities and are not general enough for the problem that we
address. \cite{Arechiga2017} is a recent work on abstraction based \emph{verification} of look-up maps.

We test the ideas proposed using a prototype tool and present its
results over case studies from the industrial processes and automotive
domain. The tool is well-integrated with the simulation-guided
falsification tool Breach \cite{breach}, and is able to use a wide
array of ranking heuristics to sort the entries in the specified
look-up map in order of importance. The tool supports arbitrary
quantitative cost functions to score executions, including the
quantitative robust satisfaction function as defined for Signal
Temporal Logic specifications in~\cite{DonzeM10,DonzeFM13}.  We
demonstrate the capabilities of the tool in localizing possible
sources of fault to look-up map entries, as well as in identifying
promising regions of parameter spaces in order to optimize the
closed-loop system performance.

\section{Problem Setting}

\noindent\textbf{Software Components and Scored Executions.} In this
work we focus on software components that occur in controllers of
dynamical systems.  For our purposes, a software component  is a
module that transforms inputs into outputs.  Let $\system$ be a
software component, and let the execution (or \emph{run}) of $\system$
given an input $z$ be denoted as $\system(z)$.  We consider a setting
where we are given a scoring  of the runs of $\system$ on a finite set
of inputs $\myZ$:  each run $\system(z)$ for $z\in \myZ$ is given a
real-valued \emph{score}, denoted $\score(\system, z)$, with negative
values being ``bad'' (denoting \emph{failed} or suboptimal runs), and positive
values being ``good'' (denoting passed or \emph{successful} runs).
This extends the boolean notion where the score set can be viewed as
$\set{-1,1}$.  Intuitively, the score of a run indicates our
satisfaction level on how good that run is; the higher the score, the
more satisfied we are with the execution.  As an example, given a
temporal logic (\eg Signal Temporal Logic~\cite{MalerN13})
specification $\stlprop$, the successful runs $\system(z)$ are those
which satisfy $\stlprop$, and the failed runs are those which do not
satisfy $\stlprop$.  Quantitative scores can be given to runs based on
quantitative semantics of the logic (\eg, the \emph{robustness}
values, ranging over reals, corresponding to a Signal Temporal Logic
formula $\stlprop$~\cite{DonzeM10,DonzeFM13}).

\smallskip \noindent\textbf{Look-up Maps.} Software
components invoke real-valued functions in their executions.  When
these functions do not have a closed form representation, or have a
closed form representation that is computationally inefficient, they
are instead stored in the form of \emph{look-up maps} as follows.  A
real-valued (finite) \emph{map} $\lutmap$ is a function from a finite
subset of $\reals^d$ to $\reals^n$.  The domain of $\lutmap$ (which is
a finite set)  is denoted by $\dom(\lutmap)$, and we refer to
elements $\entry\in \dom(\lutmap)$ as \emph{map entries}.  Let
$f_\lutmap^{\interpol}:  C \rightarrow \reals^n$ be the completion of
$\lutmap$ using some interpolation scheme $\interpol$, where $C$ is a
bounded subset of $\reals^d$.  The stored map $\lutmap$ is a discrete
representation of the function $f_\lutmap^{\interpol}$, and for a
given value $\entry$ is used to compute
$f_\lutmap^{\interpol}(\entry)$ using an interpolation scheme. We
abuse notation and use
$\interpol(\lutmap(\entry_1),\ldots,\lutmap(\entry_l))$ to denote the
result of applying the interpolation scheme $\interpol$ to the values
$\lutmap(\entry_1),\ldots,\lutmap(\entry_l)$.  The most commonly known
interpolation scheme is linear interpolation.  Controller software
components also use other interpolation schemes, such as nearest
neighbor interpolation, bilinear and bicubic interpolation
\cite{simulink}. Multivariate interpolation schemes may offer a better
approximation of function values based on the values of neighboring
points~\cite{Gasca2000}.

Given $\entry \in C$, we define $\depends\left(\lutmap, \interpol,
\entry\right)$ to be the values in the finite map domain
$\dom(\lutmap)$ which are used to compute
$f_\lutmap^{\interpol}(\entry)$.  Formally,
$\depends\left(\lutmap, \interpol, \entry \right)$ = 
\begin{equation}
\label{eq:depends}
\left\{\!\!\!
\begin{array}{ll}
    \set{\entry} 
        & \hspace*{-2mm}\text{if } \entry \in \dom(\lutmap)\\
    \set{ {\entry}_1,.., {\entry}_l} 
        & \hspace*{-2mm}\text{if }
          f_\lutmap^{\interpol}({\entry}) = 
          \interpol(\lutmap(\!{\entry}_1\!), \ldots, \lutmap(\!{\entry}_l\!)).
\end{array}
\right.
\end{equation}

Thus, $\depends\left(\lutmap, \interpol, {\entry}\right)$ indicates
which of the map domain entries $ {\entry}_1, {\entry}_2 \dots,
{\entry}_l$ are used by the interpolation scheme $\interpol$ to define
$f_\lutmap^{\interpol}\left({\entry}\right)$.  Consider a run $z$.
Given a map entry $\entry$ in the finite set $ \dom(\lutmap)$.  we say
that $z$ has  a \emph{map access for $\entry$} or $z$  \emph{accesses
$\entry$} if
(a)~there exists a query for  $f_\lutmap^{\interpol}(\entry)$ in $z$; or
(b)~there exists a query for  $f_\lutmap^{\interpol}(\entry')$ in $z$
for some $m'$ such that $\entry \in \depends\left(\lutmap, \interpol,
\entry' \right)$ (\ie the map value at $\entry$ is queried during an
interpolation in the execution $z$).

\smallskip \noindent\textbf{Parameter Localization Rankings.}
In addition to look-up maps, controller software often incorporates
several \emph{parameters} -- these are certain variable entities
that can take values in some  (usually) quantitative domain (\eg, gain constants).
Look-up maps can be viewed as a set of parameters (each $\entry \in \lutmap$ defines
a parameter with value $\lutmap(\entry)$).
The parameter localization problem is to narrow down problematic
parameters, in case of sub par controller performance,  amongst the
total set of parameters (whether the parameters are in the form of look-up map entries,
or are of other forms)  in the code.
Our approach to localization is to construct a \emph{ranking} of the parameters, in
decreasing order of our belief in them being problematic.
Our parameter  ranking approach
can be applied for (i)~\emph{repair} -- in this case we
are given a hard requirement that must be met, and executions are
scored negative iff they do not satisfy the requirement; or
(ii)~\emph{robustness and optimization} -- where we score the least
desirable runs negatively (for example, in the case of Signal Temporal
Logic where all the runs satisfy a given specification $\stlprop$, we
can shift the quantitative robustness values of runs by some negative
constant, so that executions which originally had a low positive
robustness score value get a negative score after shifting and get
into the undesired class).  

In the sequel, we fix $\system$ and $\lutmap$ and $\interpol$, and
omit them (when unnecessary)  for notational simplicity.

%

\section{Ranking based on Similarity Coefficients}

Similarity coefficients~\cite{Jones05,WongDGL14} have been widely used
in pattern analysis problems for classification and
clustering~\cite{Choi10,Jackson89,HoferPAW15,Passos16}.  A binary
similarity coefficient $\phi(X,Y)$  assigns a measure of similarity
between two binary categories $X$ and $Y$.  In our setting, $X$ is the
category of failed (or suboptimal) executions, and $Y$ is the category
of executions that are affected by an entry $\entry$.  In this
approach, we compute a ranking $\phi(X,\entry)$ for each entry
$\entry$, and sort the entries based on the $\phi(X,\entry)$ values.
In the extant approach for traditional software debugging, the
categories $X$ and $Y$ are Boolean, \ie, for a given category $X$, the
only relevant property is whether an instance belongs to $X$.  In our
case, instances are assigned a \emph{score} measuring how well they
belong to $X$.  For example, real-valued scores on executions can
correspond to a quantitative measure of how well an execution
satisfies a given logical property, by measuring the execution's
distance from the set of executions not satisfying the property. We
first extend the standard binary similarity coefficient to this
quantitative setting.  For ease of presentation, we restrict ourselves
to look-up maps. 


\subsection{Basic Quantitative Approach: Preliminaries}
\label{subsection:Basic}


Suppose that executions in $\myZ$ are scored. Further, suppose that a
run $z\in \myZ$ queries the value $f_\lutmap^{\interpol}(\entry)$ for
$m\in \dom(f_\lutmap^{\interpol})$. This value is then constructed
using the values for the entries in $\depends\left(\entry \right)$.
The execution of the run $z$ thus depends on the entries in
$\depends\left(\entry \right)$.  For $\entry\in
\dom(f_\lutmap^{\interpol})$ and $\entry'\in \dom(\lutmap)$ let
$\maffect\left(\entry, \entry' \right) = 1$ if $\entry' \in
\depends\left(\entry \right)$, and $0$ otherwise. In other words, when
a run $z$ attempts to compute $f_\lutmap^{\interpol}(\entry)$, then
for each entry $\entry' \in \dom(\lutmap)$ for which
$\maffect(\entry,\entry') = 1$, we say that $\entry'$ affects $z$.




In general, a run $z$ may query several map values
$f_\lutmap^{\interpol}(\entry_1)$, $f_\lutmap^{\interpol}(\entry_2)$,
$\dots$ during its execution.  If any of $\entry_1$, $\entry_2$,
$\dots$ are affected by a map entry $\entry'$, then $\entry'$  affects
$z$.  This is captured by the (Boolean for now) function
\begin{equation}\raffect\left(z, \entry' \right) =
\begin{cases}
1 & \text{if } z \text{ queries  some entry } m \\[-2mm]
&  \text{with }\maffect\left(\entry, \entry' \right)  = 1\\
0 & \text{otherwise}
\end{cases}
\end{equation} 

\newcommand\addSpace{\rule{0pt}{3ex}}   
\begin{table}[t]
\centering
\begin{tabular}{lll}
  \toprule
$F^\entry_A$ & = & 
$\sum_{z\st\score(z) < 0}  \raffect\left(z, \entry \right)  \cdot
   \score(z)$ \\   \addSpace

$P^\entry_A$ & = & 
$\sum_{z\st\score(z) \ge 0}  \raffect\left(z, \entry \right) \cdot
   \score(z)$ \\  \addSpace   
$F^\entry_U$ & = & 
$\sum_{z\st\score(z) < 0 ~\text{and}~\raffect(z, \entry) = 0}
   \score(z) $ \\ \addSpace
$P^\entry_U$ & = &
$\sum_{z\st\score(z) \ge 0~\text{and}~\raffect(z, \entry) = 0}
   \score(z)$ \\ \addSpace
$P$ & = & $\sum_{z\st\score(z) \ge 0}  \score(z)$ \\ \addSpace

$F$ & = & $\sum_{z\st\score(z) < 0}  \score(z) $ \\ 
\bottomrule
\vspace*{0.05em}
\end{tabular}
\caption{Building blocks for Quant. Similarity Coefficients\label{tab:rankings}}
\end{table}

We recall that the ordering of a similarly coefficient presents a
suspiciousness ranking of the indices in $\lutmap$.  A similarity
coefficient based rank  $\phi: \dom(\lutmap) \rightarrow \reals_+$
assigns a numerical value $\phi(\entry)$ to each entry $\entry\!\in\!
\dom(\lutmap)$; if the rank of a particular $\entry\!\in\! \dom(\lutmap)$
is greater than that for another entry $\entry'$, then $\entry$ is
considered more suspicious.  We define a few building blocks for
defining quantitative binary similarity coefficients in
Table~\ref{tab:rankings}, and briefly discuss these next.

\begin{enumerate}[leftmargin=1.5em]
\item 
$F^\entry_A$ is the sum of (negative) scores of the failed runs
which are \emph{affected} by entry $\entry$. The suspiciousness of
$\entry$ should increase with the value\footnote{In many applications,
\eg, when using Signal Temporal Logic and its quantitative
interpretation, a slightly negative score (which is close to $0$, \eg
$0.04$)  of an execution is a noteworthy event as this means a
property is violated.  As defined, $F^\entry_A$ however effectively
discards such runs as $\raffect\left(z, \entry \right)$ is weighted by
the score of the run.  If this is undesired, we can ``shift'' the
negative by a negative constant, and similarly the positive scores by
a positive constant in a preprocessing step so that scores with a low
absolute value do not arise.} of $\abs{F^\entry_A}$.

\item 

$P^\entry_A$ is the sum of (positive) scores of the passed runs
\emph{affected} by $\entry$.  The suspiciousness of $\entry$
should decrease for higher $P^\entry_A$.

\item 

$F^\entry_U$ is the sum of (negative) scores of the failed runs
\emph{unaffected} by entry $\entry$.  The suspiciousness of $\entry$
should decrease with an increase in $\abs{F^\entry_U}$, as a high
$\abs{F^\entry_U}$ indicates that the problem lies away from $\entry$.

\item 
$P^\entry_U$ is the sum of (positive) scores of the passing runs
\emph{unaffected} by entry $\entry$.  This quantity is not important,
as we are interested in executions which are problematic; not
executions which are problem-free and do not access $\entry$.

\item $P$ is the sum of (positive) scores of the passed runs.

\item $F$ is the sum of (negative) scores of the failed runs.

\end{enumerate}

\subsection{Basic Quantitative Similarity Coefficients}
\label{subsec:basicquantitative}

Similarity coefficients can be constructed based on the quantities
$F^\entry_A , P^\entry_A, F^\entry_U, P^\entry_U, F, P$ defined
previously.  We refer the reader to~\cite{Choi10} for $76$  similarity
coefficients that have been studied before in the Boolean case.  


In this work we focus on three illustrative similarity coefficients --
the Tarantula similarity coefficient, the Kulczynski coefficient, and
the $D^*$ similarity coefficient.

\smallskip\noindent\textbf{Tarantula similarity coefficient.} The
Tarantula tool~\cite{JonesHS02,Jones05}, given a test suite where each
test is labeled as either passing or failing, uses similarity
coefficients to rank the statements of programs in decreasing order of
suspiciousness.  The ranking is presented in the form of a visual map
with different statements getting color and brightness levels
according to the value of the similarity coefficient for that
statement.  The Tarantula similarity coefficient for potentially
faulty map entries extended to the case where instead of pass/fail
executions we have quantitative values, is given as follows:
\begin{equation}
\label{eq:tarantula}
R_{\taran}(\entry) =  \frac{{F^\entry_A}/{F}}{{F^\entry_A}/{F}\ +\ {P^\entry_A}/{P}}
\end{equation}
%
\smallskip\noindent\textbf{Kulczynski similarity coefficient.}
This similarity coefficient extended to the quantitative setting is:
\begin{equation}
\label{eq:kul}
R_{\kul}(\entry)  = \frac{\abs{F^\entry_A}}{\abs{F^\entry_U} \, +\, P^\entry_A}
\end{equation}

\smallskip\noindent\textbf{$D^*$ similarity coefficient.} The  $D^*$
coefficient~\cite{WongDGL14} is based on the Kulczynski similarity
coefficient to give more importance to failed executions which are
affected by an map entry $\entry$ compared to (i)~failed executions
which are not affected my $\entry$ (as there might be other map
indices which might be to blame for those other executions), and
(ii)~successful executions which are affected by $\entry$ (as failures
are more relevant than successes).  This adjustment is done by raising
the numerator $\abs{F^\entry_A}$ in the Kulczynski similarity
coefficient to a positive power $\gamma \ge 1$.  The
$D^*$ coefficient extended to our  quantitative setting is:
\begin{equation}
\label{eq:dstar}
R_{\dstar}^\gamma(\entry)  = \frac{\abs{F^\entry_A}^{\gamma}}{\abs{F^\entry_U} \, +\, P^\entry_A}
\end{equation}
Note that we are only interested in the order relation imposed by the
ranking functions, not the numerical values themselves.  The
work~\cite{WongDGL14} found $\gamma=2$ to be best in their
experiments.  They also  compared this coefficient to some of the
other similarity coefficients (\eg, the Tarantula coefficient), and
found the $D^*$ coefficient to be better than others (in the sense of
buggy statements being ranked closer to the top).

\subsection{Utilizing Continuity and the Metric Space Structure}
\label{subsection:Continuity}

In many cases, the map $\lutmap$ approximates a \emph{continuous}
function $f_\lutmap^{\interpol}$.  The value of
$f_\lutmap^{\interpol}(\entry)$ hence while depending directly on the
map entries in $\depends\left(\entry \right)$, is also correlated with
the map values at nearby map entries.  Consider a situation where we
have $51$ failed executions, each scored the same, where each failed
execution is affected by a different map entry (thus there are $51$
potential map entries ($\dom(\lutmap)=51$).  Of these, $50$ map
entries  are clustered very closely, and the one remaining entry is
far away from these $50$.  Intuitively then, the $50$ clustered map
entries should be viewed as more problematic than the other remaining
entry.  Additionally, design engineers are typically interested in
identifying problematic regions of a look-up map. A problematic region
in $f_\lutmap^{\interpol}$ might indicate that the engineers need to
``refine the grid'' in the map $\lutmap$ in the identified region,
rather than simply changing the map values (keeping the grid
resolution the same).


We account for correlation in suspiciousness of proximal map entries
by introducing a dependence between map entries the decays with
distance.  A modular way to accomplish this goal is to build upon the
approach of Subsection~\ref{subsection:Basic} by defining quantitative
extensions of the function $\maffect$ and $\raffect$ which account for
the fact that the values of neighboring map entries are related.  The
new similarity coefficients can then be constructed which simply use
these new $\maffect$ and $\raffect$ functions.  The quantitative
version of $\maffect$ is defined as:
\begin{equation}
\label{eq:maffectQuant}
\maffect\left(\entry, \entry' \right) =
\begin{cases}
\lambda_M^{\dist(\entry,\entry')} & \text{ if } \dist(\entry,\entry') \leq r_M\\
0 & \text{ otherwise.}
\end{cases}
\end{equation}
where $0\!<\!\lambda_M\!<\!1$ is a decay constant , and $r_M$ is a quantity denoting the
radius of influence.
The value $\maffect\left(\entry, \entry' \right) $ quantifies how much the map value at
entry  $\entry'$ affects the map value at entry $\entry$.
These quantities can in the general case be dependent on $\entry$ --
if $f_\lutmap^{\interpol}$ at $\entry$ is changing very fast, then
$\lambda_M$ and $r_M$ at $\entry$  will be small.

Using the above defined $\maffect$ function
we give a quantitative version of the function $\raffect$
which quantifies which map entries affect an execution as follows:
\begin{equation}
\label{eq:raffectMetric}
\hspace*{-3mm}
\raffect(z, \entry') = \!\!\!\!
\max_{
\begin{array}{l}
\entry\st z~\text{queries}~\entry\ \\
\wedge\ \dist(\entry,\entry') \leq r_M
\end{array}
} 
\!\!
\maffect\left(\entry, \entry' \right).
\end{equation} 

$\raffect\left(z, \entry' \right) $ can also be defined in other ways
(\eg taking a sum instead of the maximum).  We compare $\raffect$ to
the binary $\raffect$ function of Subsection~\ref{subsection:Basic} --
there $\raffect\left(z, \entry' \right) $ was either $0$ or $1$ based
on whether the map entry $\entry'$ was utilized or not during the
execution $z$.  Now, $\raffect\left(z, \entry' \right) $ gives a more
nuanced estimate of the importance of the map value for entry
$\entry'$ (accounting for the fact that neighboring map entries should
have related map values).

The new ranking coefficients corresponding to $R_{\taran}$, $
R_{\kul}$, and $R_{\dstar}^\gamma$   can be defined as in
Equations~\eqref{eq:tarantula},~\eqref{eq:kul},  and~\eqref{eq:dstar},
by plugging in the new quantitative functions $\raffect$ and
$\maffect$ in the defining equations.

\subsection{Incorporating Frequency of Access}

The rankings of the previous two subsections  do not incorporate the
\emph{frequency} of map accesses inside an execution.  Consider a
situation in which bad executions have a tendency to repeatedly access
a certain portion of the map during the course of the executions.  If
a map region is repeatedly accessed during an execution, one natural
heuristic is to give that map region more importance.  For a given run
$z$ and a map entry $\entry$, a frequency measure  of accessing a
particular map region around $\entry$ can be expressed by generalizing
the $\raffect$ function as follows.  Let $|z|$ denote the number of
map queries in $z$, that is, if in $z$ we have the (possibly
non-distinct) queries $f_\lutmap^{\interpol}(\entry_1),
f_\lutmap^{\interpol}(\entry_2), \dots
f_\lutmap^{\interpol}(\entry_p)$ then $|z|$ is $p$.  The $k$-th  query
in $z$ is denoted by $z[k]$.
The function $\fraffect$ generalizes $\raffect$, and is defined as
\begin{equation}
\fraffect\left(z, \entry \right) =
\frac{
\sum_{k=1}^{|z|}
\maffect\left(z[k], \entry \right)}
{|z|}
\end{equation}
for  $\entry\in \dom(M)$.

Intuitively, $\fraffect\left(z, \entry \right) $ can be thought of as
a weighted fraction corresponding to the effect of the region around
$\entry$ on the execution $z$.  The function $\fraffect$ can be
defined using the binary $\maffect$ function; or the quantitative
$\maffect$ function of Equation~\eqref{eq:maffectQuant} incorporating
entry correlation.  As an example, if we use the binary $\maffect$
function, and $\fraffect\left(z, \entry \right) $ is $0.4$, then it
means that $40\%$ of the function calls $
f_\lutmap^{\interpol}(\cdot)$ during the course of $z$ were affected
by $\entry$.  Analogues to
Equations~\eqref{eq:tarantula},~\eqref{eq:kul}, and~\eqref{eq:dstar}
can be obtained by replacing $\raffect$ by $\fraffect$; and letting $P
= P_U^\entry + P_A^\entry$, and $F = F_U^\entry + F_A^\entry$.

\section{Set Spectra Based Methods}

An alternative to the binary coefficient based ranking method is to
define various \emph{sets} of map indices for inferring possible
problematic values.  The work in~\cite{Pan92heuristicsfor} defines
several  sets of interest in software fault localization  using set
algebra operations.  We adapt the union model, which is 
the most promising set-spectra based heuristic, to
our quantitative setting of map lookups.



The union model~\cite{Pan92heuristicsfor,AgrawalHLW95} for software
fault localization -- given a failed run -- looks at statements that
are executed in the buggy execution, but not in \emph{any} successful
run. The intuition behind the model is that if certain statements are
executed \emph{only} in failing runs, those statements are likely to
be  buggy.  The term union comes from the fact that under this
heuristic, the set of suspicious statements is given by 
\[
\displaystyle\bigcup_{\substack{\text{statement }\mathsf{s} \text{ executed by } z\\
 \text{ s.t. } \score(z) <0} }
\set{\mathsf{s}} \qquad\qquad  
\setminus \quad
\displaystyle\bigcup_{\substack{\text{statement } \mathsf{s} \text{ executed by } z\\
 \text{ s.t. } \score(z) > 0}} 
\set{\mathsf{s}}
\]

The work in~\cite{RenierisR03} found in their experiments that buggy
statements often are executed in successful runs, and so the union
model in many cases fails to label these buggy statements as
suspicious (\ie, the above set is empty); however, in cases where the
union model \emph{does} label statements as suspicious, the labelled
statements are almost always buggy. That is, the union model has a
very low false positive rate in labelling buggy statements (and a high
false negative rate). A very low false positive rate is extremely
attractive in bug finding, thus, we chose to explore the performance
of the union method for our look-up map setting.  In case the union
method labels every region as non-buggy, we can fall back on other
methods, \eg, the methods of the previous section.

\mypara{Utilizing Continuity of Map Values}
We modify the union model in this case as follows.  We ask for the
following (at a high level):  is there a map access $m$ in a run $z$
with a negative score $\score(z)$, such that  all positively scored
runs $z'$ access map indices  at least $r_M$ distance away from $m$?
That is, for the set of executions $\myZ$, is there an area in
$\dom(f_M^{\interpol})$ that is (i)~accessed only by failing runs, and
(ii)~is at least $r_M$ distance away from the areas accessed by
successful runs?

Formally, let $\access(n,m)$ be the predicate on whether a run $z$
access map entry $m$,  let $M_F$ be the set of map entries  accessed by  faulty
(negatively scored) runs, and let $M_S$ be the set of map entries
accessed by successful (non-negatively scored) runs.  That is,
\begin{compactitem}
\item 
$
M_F = \left\{
m \left\lvert
 \exists\, z \st \score(z)\! <\! 0 \wedge \access(z,m) = \true
\right.\right\};$ 
\item 
$
M_S = \left\{
m \left\lvert 
\exists\, z \st \score(z) \geq  0  \wedge \access(z,m) = \true
\right.\right\}$.
\end{compactitem}

For $X$ a set of map entries and $r$ a positive real number, let
$\ball(X, r)$ denote the set $\set{x'\in \dom(M)  \mid \text{ exists }
x\in X$  such that $ \dist(x, x') \leq r}$.  The set of
suspicious map indices  $\sus_U$ is defined to be:
\begin{equation}
\label{eq:Union}
\sus_U = M_F \setminus \ball(M_S, r_M).
\end{equation}

The elements in $\sus_U$ are the suspicious map entries.  In most
cases, a ranking on $\sus_U$ will not be necessary as $\sus_U$ will be
a small set, and the entire set can be classified as suspicious.  If
desired we can rank the entries in $\sus_U$ as follows.  For a map
entry $m \in \sus_U$, the quantification of suspiciousness of $m$
depends on:
(a)~the negative scores of the failed runs, and 
(b)~the distance from the map accesses in the positively scored runs.  
For a map entry $m \in \sus_U$, we call the first quantity accessed by
a negatively scored run as $s_U(m)$ and define it as follows:
\begin{equation}
  s_U(m) = \min\left\{ \abs{\score(z)}  \ \left\lvert\
  \access(\!z,m\!)\! =\! \true \wedge \score(\!z\!) \!<\! 0\right.\right\}
\end{equation}
In other words, $s_U(m) $ tells us what is the absolute value score of
the best among failing runs (\ie the score which is closest to $0$)
which is affected by $m$.  The higher the value of $s_U(m) $, the more
suspicious should $m$ be.  The second quantity for an entry $m\in
\sus_U$ is denoted $d_U(m)$ and we define it as follows:
\begin{equation*}
\hspace*{-3mm}
d_U(\!m\!) =
\min\!\left\{\!\dist(m, m') \ \left\lvert
\begin{array}{l}
\exists \, z' \st 
 \score(z') \geq 0 \,\wedge \\
 \access(\!z', m'\!)= \true \wedge \dist(m, m')\! >\! r_M
 \end{array}\right.\!\!\!\!\!
 \right\}
\end{equation*}

Note that the above value is equivalent to minimize the distance
$
\left\{\dist(m, m') \ \left\lvert\ 
 m'\in M_S\right.\right\} 
$.

In case the set in the above equation is empty, \ie,
if there does not exist a run $z'$ such that $\score(z') \geq 0$,
we let $d_U(m)$ be a high constant.
Note that since $m\in  \sus_U$, if there exists a  run $z'$ such that $\score(z') \geq 0$,
then $z'$ access $m'$ with  $\dist(m, m') > r_M$.
The quantity  $d_U(m)$ gives the separation distance between $m$ and
$M_S$ (the set of entries affected by successful runs).
The higher the value of $d_U(m) $, the more suspicious should $m$ be.
The set union based ranking  $R_U$ is a function of $d_U$ and $s_U$:
\[
R_U(m) = s_U(m) \cdot d_U(m).
\]

Note that setting $r_M = 0$ makes $R_U(m)$ be $0$ iff there exists a positively scored
run which accesses $m$ (like in the original union model).

The above method marks only map entries which are accessed, as opposed to the 
rankings of the previous section which may mark entries  that are not accessed, but lie
in suspicious areas of the $f_M^{\interpol}$ areas.
When using this method in practice, the user thus must look at the surrounding entries
in case an index $m$ is found to be buggy.

We keep only the set algebraic version of Equation~\eqref{eq:Union} for this model,
and not define quantitative extensions  based on smoothness of $f_M^{\interpol}$
(as in Subsection~\ref{subsection:Continuity}) because this heuristic is based on set-algebra. We expect a high percentage of entries in  $\sus_U$ to be problematic
(the set $\sus_U$ is expected to be small), thus further
quantitative refinements are not of much use.

\section{\hspace*{-1mm}Evaluating Effectiveness of Proposed Approach\hspace*{-10mm}}

For fault localization in traditional software,
the EXAM score~\cite{WongGLAW16} is a commonly used measure to quantify the
effectiveness of the different  techniques. 
In our context, it can be defined as 
{\small
\[\examscore=
\frac{\mspace{-5mu}
\begin{array}{c}
\text{number of look-up map entries examined}\\
\text{(in decreasing order of their scores)}\\
\text{till an entry deemed problematic is encountered}
\mspace{-10mu}
\end{array}
}{
\text{total number of look-up map entries}}
\ \times 100
\]
}
A lower score is better as it indicates that only a small number of entries are ranked above the
``true'' buggy ones.
However, in engineering practice, a percentile relative score is not always sufficient~\cite{Parnin2011};
thus we also define an \emph{absolute} version of the EXAM score as 
{\small
\[\absexamscore=\mspace{-10mu}
\begin{array}{c}
\text{number of look-up map entries examined}\\
\text{(in decreasing order of their scores)}\\
\text{till an entry deemed problematic is encountered}
\end{array}
\]
}

In the look-up map context, $\examscore$, and  $\absexamscore$ values are not 
always  appropriate metrics.
While  these scores  are  relevant when the maps contain isolated entries that may be
the cause of a fault, in typical practice, a \emph{region} of  the map (which
often correspond to regions of operation of the plant and the controller)
contains suboptimal entries.
In essence, we want to \emph{cluster} map entries based on their rank values, and present these clusters to the user in order to identify suboptimal map regions.
Ideally, there should be a small number of clusters containing the
high-ranked map entries, and moreover each such cluster containing high-ranked map entries
should contain minimal  \emph{low}-ranked entries.
There is a plethora of cluster analysis algorithms and tools~\cite{Aggarwal2013}. 
For one or two-dimensional look-up tables, one of the simplest methods is by visual inspection
of a \emph{heat-map} of the ranked entries (Matlab provides a heat-map visualization 
function).
Our tool incorporates this heat-map visualization functionality.


\section{Case Studies}


In this section, we empirically evaluate the efficacy of  the ranking heuristics on
several case studies.
In Subsection~\ref{subsec:BasicCaseStudies} we investigate the ranking heuristics
on smaller examples, and in Subsection~\ref{subsec:IndustrialCaseStudies} we
run the ranking methods on industrial case studies.
%
%
%

In this section, we use the term look-up table (\lut) for readers
 familiar with the eponymous \simulink block for implementing
 $N$-dimensional look-up.  Understanding how each ranking heuristic
 performs on toy models, can help provide CPS designers with
 guidelines on choosing the right ranking heuristic based on their
 design-type.  We assume reader familiarity with temporal logics such
 as Signal Temporal Logic (STL).  We refer the readers
 to~\cite{MalerN13,DonzeFM13} for STL syntax, and semantics (boolean
 and quantitative).


\subsection{Basic Models}
\label{subsec:BasicCaseStudies} 

\subsubsection{\textbf{Nonlinearity Cancellator}}
We designed a toy model $\toymodel$ to mimic canceling a nonlinearity
in the input $u(t)$ in \simulink. $\toymodel$ has two outputs
functions:
\begin{equation}
\hspace*{-2mm}
\begin{array}{l@{\hspace{2em}}l}
y_1(t) = u(t) \cdot M(u(t)) &
\displaystyle y_2(t) = \sum_{k=1}^{\lfloor \frac{t}{\Delta} \rfloor} \Delta\cdot y_1(k\Delta) 
\end{array}
\end{equation}

In the equation for $y_2(t)$, $\Delta$ represents the fixed time step
used for simulating $\toymodel$, ($\Delta = 0.1$ sec. for this
experiment).  In the equation for $y_1(t)$, $M(x)$ represents a \lut
representing the nonlinear function $\frac{1}{x}$.  In our example, we
use 
\begin{equation}
\label{eq:domM}
\dom(M) = \setof{e \mid e = 0.1\cdot z \wedge z\in \nat \wedge z \in [1,90]},
\end{equation}
(thus there are $90$ entries in $M$), and for each $e$ in $\dom(M)$, 
we let $M(e) =
0.01\cdot\mathsf{round}(\frac{100}{e})$, \ie, the approximation of
$\frac{1}{e}$ up to two decimal places\footnote{Note
$\mathsf{round}(a)$ rounds $a$ to the nearest integer.}.

The input signal $u(t)$ is specified in terms of $11$ equally spaced
control inputs between times $0$ and $30$ secs, where $u(t)$ is in
$[0.09,9.01]$ at each control point, and is a linear interpolation of
the values at control points for all other times. The domain of the
input signal is purposely chosen to exceed the domain of the LUT $M$
to exercise the extrapolation of values performed by the LUT\footnote{
We note that the semantics of LUTs in \simulink allow for values
outside the domain of the LUT to be input to the LUT. The LUT output
computed is then obtained by linear extrapolation.}. The range of
$u(t)$ and the chosen $M$ together guarantee that $\forall u(t) \in
[0.09,9.01]$, $u(t)\cdot M(u(t)) < 1.4$.  
We are interested in checking the model against the following STL
requirements:
\begin{equation}
\hspace*{-6mm}
\begin{array}{l@{\hspace{1em}}l}
\ftoyone \triangleq \alw_{[10,30]}(|y_1\! -\! 1| \!<\! 0.4) &
\ftoytwo \triangleq \alw_{[0,30]}(y_2 \le 30) \\
\end{array}
\end{equation}
We then seed the \lut $M$ with a bug, by changing $M(2)$ to $0.8$  (original value $0.5$).  We
excite the model with $100$ randomly chosen piecewise linear signals
$u(t)$, 
and then evaluate the ranking on entries produced by each of the
heuristics.

\noindent\emph{Results.}
The top $3$ entries deemed most important by each of the
heuristics are reported in Table~\ref{tab:table_toy12}. 
\begin{table*}[t]
\centering
\begin{tabular*}{.99\textwidth}{@{\extracolsep{\fill}}l c c c l}
\toprule
\vspace{0.3em}%
Ranking              & \muctwo{Top $3$ entries for $\toymodel$ ($1$D LUT, size $90$) }  
                     & \muctwo{Top $3$ entries for $\nonlmodel$ ($2$D LUT, size $1681$)} \\
\cline{2-3} \cline{4-5}
Heuristic            & Req. $\ftoyone$ & Req. $\ftoytwo$ & Req. $\fff$ &  $\absexamscore$ \\
& \\
\midrule                     
$R_{\taran}$           & --       & --           & (-9.5,-0.5) , (-9.5,0.0) , (0.0,-0.5)  & -- \\
$R_{\taran}$ (metric)  & --       & --              & (-10.0,7.5) , (-10.0,8.0), (-10.0,8.5) & 1  \\ 
$R_{\dstar}^2$         & 1.7,1.4,1.1 & 2.0,1.7,2.3  &   (-10.0,0.5) , (-10.0,1.0), (-10.0,4.5) & $>$ 3\\ 
$R_{\kul}$ (freq.)     & 2.0,1.7,1.9 & 2.0,1.7,2.3   &   (-9.5,0.5)  , (-9.5,1.0) , (-10.0,1.5) & $>$ 3\\ 
$R_{\dstar}^2$ (freq.) & 2.0,1.9,2.3 & 1.9,1.8,1.6  & (-10.0,0.5) , (-10.0,1.0), (-10.0,4.5) & $>$ 3\\
$R_U$                  & --       & --           & (-10.0,10.0), (-10.0,9.0), (-10.0,9.5) & 1  \\
\bottomrule
\vspace{0.1em}%
\end{tabular*}
\caption{The most significant $3$ entries as deemed by the ranking
heuristics for experiments on the models in~\ref{subsec:BasicCaseStudies}
 \label{tab:table_toy12}}
\end{table*}
%

We get no information from the $R_{\taran}$ ranking heuristic as all
entries are weighted equally.  For the first property ($\varphi_1$), the
ranking heuristics that take the frequency of an entry into account perform
generally better than the heuristic based on a Boolean notion of access.
This is expected as every violation of the requirement (which corresponds
to accessing the faulty region of the \lut), has an additive effect on the
frequency-based ranking heuristics. The union spectrum based methods are
also ineffective because the way we designed the experiment, all \lut
entries are accessed by each trace. This is achieved by supplying an input
signal with the first linear segment that is a ramp from $0.09$ to $9.01$
within the first $3$ seconds. As this example had a seeded bug, we can
compute the $\absexamscore$. This score is $-, -, 2, 1, 2, -$ corresponding to
the six ranking functions in Table~\ref{tab:table_toy12} (here we take the worst
scores corresponding to the two specification functions $\ftoyone$ and$\ftoytwo$).
Recall that the union spectrum method works well when there are
certain \lut entries accessed only by the failing traces.

\subsubsection{\textbf{Two-Dimensional Nonlinear System}}
In this experiment, we designed $\nonlmodel$, a model to represent the
\textbf{F}eed\textbf{F}orward control of a nonlinear dynamical system.
The plant is a $2$-dimensional nonlinear (unstable) system with
dynamics described by ODEs on the left side of Eqn.~\eqref{eq:ff}.
\begin{equation}
\label{eq:ff}
\hspace*{-7mm}
\begin{array}{l@{\hspace{1em}}l}
\begin{array}{l}
\dot{x_1} = -3x_1 + 2x_1x_2^2 + u\\
\dot{x_2} = -x_2^3 - x_2 
\end{array} &
\fff \triangleq \alw_{[0.8,2]} (|x_1| < 0.8)
\end{array}
\end{equation}

The control action $u$ used is one which trivially cancels out the
nonlinearity in the first state's dynamical equation ($u = -2 x_1
x_2^2$).  We observe that once $u$ cancels out the $2x_1x_2^2$ term,
the rest of the system is trivially asymptotically stable\footnote{The
Lie derivative of the resulting system with $V(x_1,x_2)$ as the
Lyapunov function is $-(3x_1^2 + x_2^4 + x_2^2)$, and as the term
inside the parentheses is a sum-of-squares polynomial, the Lie
derivative of $V(x_1,x_2)$ is negative everywhere.} with the
$V(x_1,x_2) = x_1^2 + x_2^2$ as a Lyapunov function certifying
stability.  This guarantees that for any $c$, the system never leaves
the set $V(x_1,x_2) < c$. 

We note that the computed feedforward control action is a polynomial
in the plant state that can be represented using a $2$-dimensional
\lut in which we add an entry at intervals of $0.5$ for integer values
of $x_1$ and $x_2$ in the range $[-10,10]$. This gives a \lut with
$41^2=1681$ entries.  For the purpose of simulating the system, we
pick a random initial state $x_1(0) \in [-10,0]$ and $x_2(0) \in
[0,10]$.  We then introduce bugs in $30$ of the entries that
correspond to $x_1$ in $[-10,-8]$, and $x_2$ in $[7.5,10]$. The bug
basically multiplies each entry by $-2$. While the original system is
globally asymptotically stable, observe that the bug may cause
unstable behavior in the system.


The STL requirement on the right side of Eqn.~\eqref{eq:ff} relates to
the settling time of $x_1(t)$. We run $100$ randomly chosen
simulations to excite the model and apply our ranking heuristics to
identify the entries likeliest to be the root cause of settling time
violations.  

\noindent\emph{Results.}
The top $3$ entries deemed most important by each of the
heuristics are reported in Table~\ref{tab:table_toy12}. 
The ranking heuristics based on quantitative similarity
coefficients do not work well for this example. (There is an
exception: the Tarantula rankings utilizing the metric space
interpretation of the \lut).  This is so because when a wrong \lut
entry is accessed in computing the feedback law, the controlled signal
deviates from the desired reference value, and all subsequent accesses
of the \lut are indexed by the deviated values.  The set spectrum
method based on the union model, on the other hand, focuses on entries
that are solely accessed by the failing traces, and does not get
misled by a frequentist reasoning approach.  The Tarantula-based
binary similarity coefficients also focus on entries appearing in
failing traces; however, unless additional weighting is provided to
the failing entries by considering nearby entries as potential sources
of failure, the Tarantula rankings are not effective. This also shows
the value of using the metric space interpretation of \lut entries.

\subsection{Industrial Case Studies}
\label{subsec:IndustrialCaseStudies}

\begin{table*}[t]
\centering
\begin{tabular*}{.99\textwidth}{@{\extracolsep{\fill}}l c c c}
\toprule
\vspace{0.3em}%
Ranking               & Top $3$ entries for $\gcstrmodel$ ($1$D LUT,  size $8$)
                      & \muctwo{Top $3$ entries for $\afcmodel$ ($1$D LUT,  size $11$)}  \\
\cline{2-2} \cline{3-4}
Heuristic              & Req. $\fcstr$  
                       & Req.  $\fafcovershoot$
                       & Req. $\fafcsettle$ \\
\midrule                     
$R_{\taran}$           & 3,2,4    &  3250,3000,2750   &  3250,2750,3000 \\
$R_{\dstar}^2$         & 4,3,5   &  2500,2250,2000   &  2250,2000,2500   \\
$R_{\kul}$ (freq.)     & 4,3,5   &  2500,2250,2000   &  2250,2000,2500   \\
$R_{\dstar}^2$ (freq.) & 4,3,5   &  2500,2250,2000   &  2250,2000,2500   \\
$R_U$                  & --      &  --      &  -- \\
\bottomrule
\vspace{0.1em}%
\end{tabular*}
\caption{Column entries contain the top $3$ entries deemed most
important by each ranking heuristic. The second column contains
entries for the continuously stirred tank reactor model. The
third \& fourth columns contain results for air-fuel ratio control
system.\hspace*{-10mm} \label{tab:cstr_afc}}
\end{table*}

\subsubsection{\textbf{Debugging a Gain-Scheduled Control System}}
 In this case
study, we look at $\gcstrmodel$, a model of a continuously stirred
tank reactor (CSTR), a common chemical system used in the industry.  A
\simulink model of this process is available as a demonstration
example from the Mathworks \cite{cstr}. The control objective is to
ensure that the concentration of the reagent in the tank is maintained
at a specified set-point. The model assumes that the controller can
gets sensor readings of the residual concentration of the reagent in
the tank, and is able to change the temperature of the coolant in
reactor's cooling jacket to control the reaction.  The control task is
complex as the process dynamics are nonlinear and vary substantially
with concentration. Hence, the system uses a Proportional + Integral +
Derivative (PID) controller that is gain-scheduled, \ie, uses
different P+I+D gains for different reference set-points for the
concentration. There are $8$ different control regimes and $3$
different lookup tables for the P, I and D gains respectively. 

In this experiment, we introduce a bug in the model. We reverse the
polarity of the P gain for the control regime corresponding to a
concentration of $3$ units (entry $M(3)$ in the P gain \lut.
Outwardly, the bug is egregious; reversing the polarity of the P gain
makes the closed loop system in that control regime unstable. However,
once the system transitions to a different concentration regime,
corrective feedback takes over, and the system eventually settles
(possibly with a longer settling time).  The STL requirement in
\eqref{eq:cstr_settles} captures that: (1) the system is excited by a
step change in the reference at $2$ seconds (enforced by the
time-interval for the outermost $\alw$ operator), and (2) the maximum
percentage deviation in the concentration signal $c(t)$ from the given
settling region is less than $2\%$ of the reference $r(t)$ after the
given settling time deadline of 2.5 seconds.
\begin{equation}
\label{eq:cstr_settles}
\hspace*{-3mm}
\fcstr \triangleq \alw_{[2,20]}\!\!\left(\!\step(\!r\!) \Rightarrow
    \alw_{[2.5,9.9]} \left(\left|\frac{c - r}{r}\right| \!<\!
    0.02\!\right)\!\!\right)
\end{equation}
We ran $100$ randomly chosen simulations and ranked entries
that are the likeliest causes for failure of requirement $\fcstr$.

\noindent\emph{Results.}
 The
results are shown in Table~\ref{tab:cstr_afc}.  
All the similarity coefficients except $R_U$ are able to find the seeded bug.
In this case, the union spectrum
based heuristic is not useful as there is no clear separation between
entries accessed by failing and passing traces. 

\subsubsection{\textbf{Identifying Sensitive Regions in an Observer}}
In this case
study, we consider $\afcmodel$, a model \cite{arch14_benchmark} for
regulating the air to fuel ratio (denoted $\lambda$) in the mixture
that undergoes combustion in gasoline engines.  As the peak
efficiency of a catalytic converter to reduce noxious emissions in the
exhaust is reached when $\lambda$ is $14.7$, this is an important
control problem.  $\afcmodel$ has a controller that uses an observer
to estimate the amount of fuel that puddles on the injection port and
is thus not used for combustion in the engine.  The observer is based
on the Aquino model for fuel puddling in the controller
\cite{arch14_benchmark}.  This observer uses two \luts for
predicting the deposit ratio and residual ratio of the fuel from
engine speed. Both \luts have the same size and are indexed by the
same quantity (engine speed), so the access patterns for entries of
both \luts are identical, and the entries shown in the results
correspond to the corresponding entries for the same engine speed.

Let $\mu(t) = \frac{\lambda-14.7}{14.7}$ and let $\theta(t)$ denote
the throttle input signal from the user. Then, we wish to identify the
regions of the observer \luts that have the most impact on the maximum
overshoot ($5\%$) and settling time ($1$ sec.) requirements on
$\lambda$ as shown in
Eqns.~\eqref{eq:overshootafc},\eqref{eq:settlingafc}.
\begin{eqnarray}
\fafcovershoot \triangleq & 
\mspace{-10mu}
\alw_{[2.5,10]} \left(\step(\theta) \Rightarrow \alw (\mu < 0.05)\right)
\label{eq:overshootafc} \\
\fafcsettle \triangleq & 
\mspace{-10mu}
\alw_{[2.5,10]} \left(\step(\theta) \Rightarrow
                 \alw_{[1,\infty]}(|\mu| < 0.01)\right)
\mspace{20mu}
\label{eq:settlingafc}
\end{eqnarray}

\noindent\emph{Results.}
Table~\ref{tab:cstr_afc} indicates that the observer region
corresponding to high engine speed does not predict the state of the
deposit ratio and residual ratio (two quantities to quantify the
amount of fuel that puddles) accurately, causing the model to behave
poorly at high engine speed.  In \cite{arch14_benchmark}, the authors
indicate that the observer was designed based on a linear model at an
operating point of $1000$ rpm.  Thus, the result above confirms the
designer hypothesis that the observer performance is poor at high
engine speeds.

\subsubsection{\textbf{Parameter Space Optimization for a Suspension Control System}}
In \cite{quartercar_simulink_control_demo}, the authors present
$\suspmodel$, a quarter car  model of an automotive suspension system
implemented in \simulink. $\suspmodel$ considers only one of the wheels
and simplifies the dynamics to that of a suspension mass suspended by
two spring-damper systems.  The controller uses a 
Proportional+Integral+Derivative (PID) scheme to provide active
assistance to the suspension system. The design objective for this
system is that the distance between the suspension mass and the
vehicle body (denoted $y$) shows acceptable transient behavior, as
sustained oscillations or a slow settling time on $y$ are a cause
discomfort to the vehicle occupants.

Let the P, I and D gains be respectively $K_p$,$K_i$ and $K_d$. Let
$K$ = $(K_p,K_i,K_d)$.  We optimize the controller performance by
looking for regions in $K$-space that correlate well with good
controller performance. In this model, there are no explicit \luts,
but we impose a grid on $K$-space and then use our tool to find grid
regions with a high rank with respect to the design objective. In
contrast to other case studies, where we use ranking heuristics
(all rankings except $R_{\taran}$  had similar behaviour, so we
picked a typical ranking function)  to
identify the cause of undesirable behavior, in this case, we use it to
search for desirable behavior. As $K$-space does not have an explicit
(user-defined) structure, this case study allows us a flexible way of
exploring the (possibly nonlinear) parameter space. We explore a
simple scheme, where we first search the $K$-space using a coarse
grid, and then impose a finer grid on the top grid element found in
the first iteration.  In each iteration, we pick $200$ simulations for
randomly chosen values of the gains within the chosen paramter
regions.  This example requires large numbers for the $K_p$, $K_i$,
$K_d$ gains, so for brevity in presentation, we assume that each of
the gains in the following discussion are $\times 10^6$.

For the first iteration, we assume that $K_p \in [100, 500]$, $K_i \in
[8, 600]$, and $K_d \in [0, 3]$.  We use $21$ equally spaced grid
points in these intervals giving rise to a total of $9261$ grid
elements.  In the first iteration, the tool identifies the grid
element $(16,18,2)$ as the highest ranked grid point according to a
simple majority of the rankings.  This corresponds to the center of
the region: $K_p \in [380,420]$, $K_i \in [481.6,540.8]$, $K_d \in
[0,0.3]$.  We then impose a finer grid on this region, with $21$
equally spaced points for each gain-parameter. After running the
second iteration, we identified  grid element $(13,14,2)$ as the top
ranked element in the new grid. This corresponds to the region $K_p
\in [402,406]$, $K_i \in [517.12,523.04]$, and $K_d \in [0,0.03]$.
Essentially, this is a narrow region for $K_p, K_i, K_d$ that
correlates well with good controller performance (i.e. low settling
time). This demonstrates the power of the tool for parameter
optimization.

\subsubsection{\textbf{Model Predictive Control of a Diesel Engine Air Path}}
Next, we
consider $\dieselmodel$, an early prototype closed-loop \simulink
model from \cite{Huang2016} of a \textbf{D}iesel engine \textbf{A}ir
\textbf{P}ath controller.  $\dieselmodel$ has a high fidelity plant
model of the air path dynamics and a model predictive controller (MPC)
to regulate the intake manifold pressure and the exhaust gas
recirculation (EGR) flow rate.  A notable feature of $\dieselmodel$ is
its scale: it has more than $3000$ \simulink blocks and over $20$
multi-dimensional lookup tables.  $\dieselmodel$ has two inputs: (1)
the fuel injection rate (denoted $\fuelrate$ and excited by a single
step of magnitude in a given range), and the engine speed (denoted
$\engspeed$, picked from a given range, but held constant during any
single simulation).  There are two outputs of interest: the intake
manifold pressure (denoted $p$) and the EGR flow rate (denoted
$\egr$). The control designers for $\dieselmodel$ indicated to us that
they are interested in two requirements:
Requirement~\eqref{eq:dieselreq1} characterizes the overshoot in the
intake manifold pressure.  Requirement~\eqref{eq:dieselreq2} relates
to how well the MPC scheme tracks the $\egr$ signal against the $\egr$
reference signal (denoted $\egr_\mathrm{ref}$).  Let $\mu = \frac{\egr
- \egr_\mathrm{ref}}{\egr_\mathrm{ref}}$. 
\begin{eqnarray}
\hspace*{-5mm}
\fdieselovershoot \triangleq & 
\alw_{[2,10]} (\step(fr) \Rightarrow \alw_{[0,10]} (p < \pmax)) 
\label{eq:dieselreq1} \\
\fdieselsettle  \triangleq & 
\alw_{[2,10]} (\step(fr) \Rightarrow \alw_{[\tau,10]}(|\mu| < v)) 
\label{eq:dieselreq2}
\end{eqnarray}
As this model is proprietary, we suppress the numeric values for the
settling time ($\tau$), the settling region ($v$), and the maximum
allowed overshoot ($\pmax$). For the \lut entries, we scale the actual
values to representative integer values.  We pick an \lut identified
as the most important for analysis by the control designer. This is a 2D
$20\times15$ \lut.
\begin{table*}[t]
\centering
\begin{tabular*}{0.99\textwidth}{@{\extracolsep{\fill}}l c c}
\toprule
\vspace{0.3em}%
Ranking              & \muctwo{Top $3$ entries for $\dieselmodel$ ($2$D LUT, size $300$)} \\
\cline{2-3}
Heuristic            & Req. $\fdieselovershoot$ 
                     & Req. $\fdieselsettle$ \\
\midrule
$R_{\dstar}^2$         & (2,9),(1,9),(2,10)  & (2,11),(1,11),(4,11)  \\
$R_{\kul}$ (freq.)     & (2,9),(1,9),(2,10)  & (2,11),(1,11),(2,10)  \\
$R_{\dstar}^2$ (freq.) & (2,9),(1,9),(2,10)  & (2,11),(1,11),(2,12)  \\
$R_{\dstar}^2$ (freq.+metric) 
                       & (2,9),(5,9),(1,8)   & (1,10),(2,11),(3,10) \\
$R_U$                  & --                  & (12,1),(10,10),(11,10) \\
\bottomrule
\vspace{0.1em}%
\end{tabular*}
\caption{Top $3$ entries deemed the most important by different ranking
heuristics for Model predictive of a Diesel Engine Airpath.
model.\label{tab:tablediesel}}
\end{table*}

\noindent\emph{Results.}
We show the results obtained
by applying our tool (for a typical similarity coefficient ranking) on $500$ random simulation runs of $\dieselmodel$
in Table~\ref{tab:tablediesel}.  For this particular experiment, the
\lut entries deemed most important have a direct interpretation as the
inputs to the \lut are model inputs $\fuelrate$ and $\engspeed$.
Thus, each entry in the \lut corresponds to a range of $\fuelrate$ and
$\engspeed$ values. We remark that the entry $(2,9)$ corresponds to a
low fuel injection rate and higher engine speed scenario, while the
entry $(12,1)$ corresponds to a higher fuel injection rate and very
low engine speed scenario. We also remark that for the settling time
requirement, we find entries accessed only by the failing
trajectories.  Interestingly, the top three entries, although from
different regions of the \lut, all correspond to boundary (edge) cases
of the \lut. This indicates that the MPC-based controller has poorer
performance w.r.t.  the settling time requirements at boundary
conditions. 
\begin{figure*}[t]
\centering
\subfigure[Importance of \lut entries for the requirement related to
overshoot on the intake manifold pressure during up-steps in fuel
injection. The results shown are w.r.t. the $R_{\dstar}$ heuristic
incorporating frequency of access and metric-space interpretation of
the table.]{%
    \label{fig:overshoot_heatmap}
    \includegraphics[height=50mm,trim={0cm 5cm 0 5cm},clip]{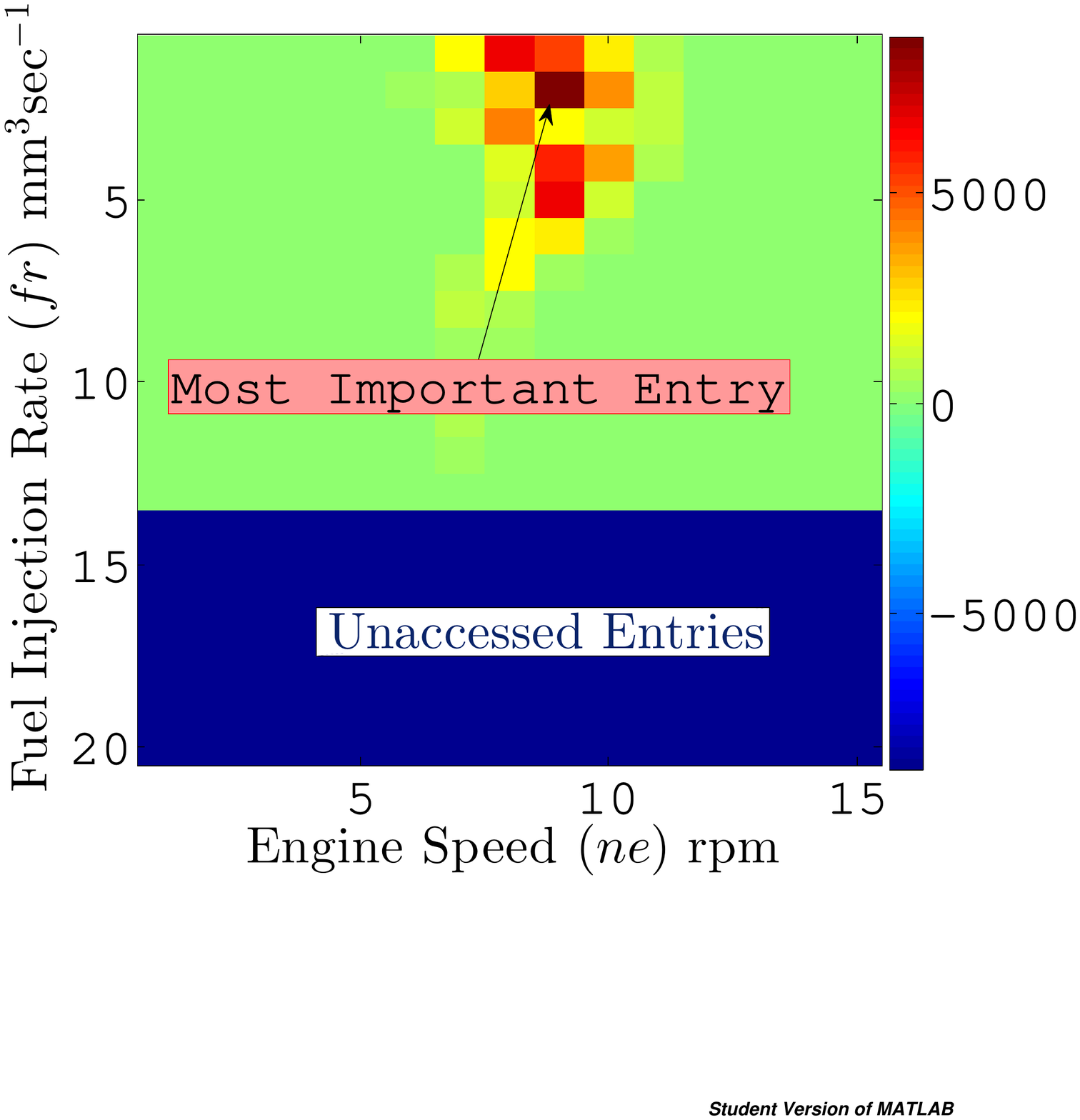}
} \qquad
\subfigure[Importance of \lut entries for the requirement related to
settling time on the EGR rate on down-steps in fuel injection. The
results shown are w.r.t. the $R_{\dstar}$ heuristic incorporating
Boolean access of entries and a metric-space interpretation of the
table.]{%
    \label{fig:settling_heatmap}
    \includegraphics[height=50mm,trim={0cm 5cm 0 5cm},clip]{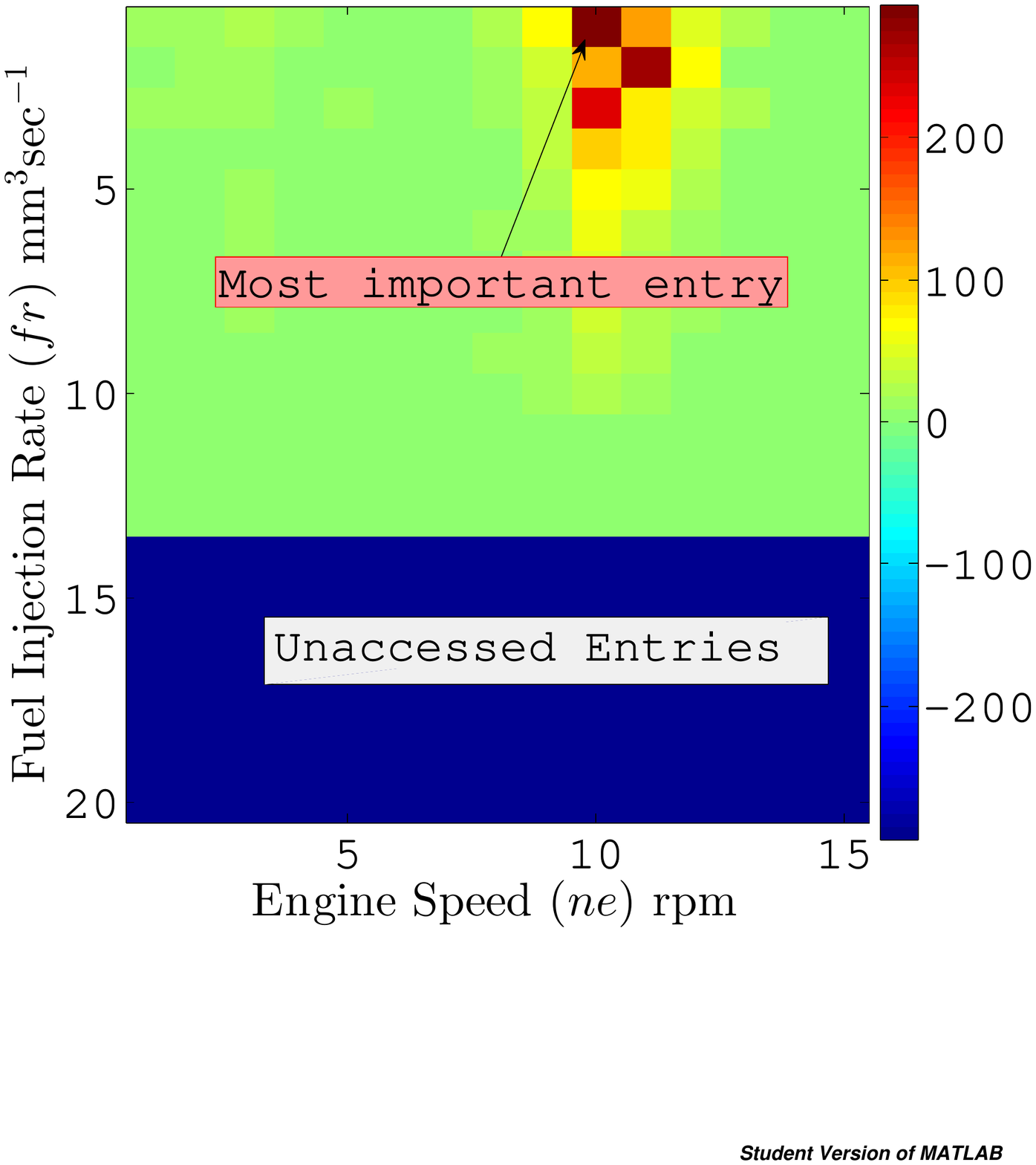}
}
\caption{Heatmap representation for \lut entries for the
$\dieselmodel$ model. Entries in blue are not accessed by any trace.
The color spectrum from light green (least) to dark red (most)
indicates importance of the entry. The numbers on the color-bar are
the rank values, that can be used to have a quantitative
interpretation of the relative importance of entries. Observe that the
graphical depiction may allow a broader judgement regarding
problematic regions in the \lut.
\label{fig:heatmap}}
\vspace*{-3mm}
\end{figure*}

In Fig.~\ref{fig:heatmap}, we present the results in the form of a heat-map that
pinpoints hot-spots in the chosen look-up table w.r.t. a given requirement.  The
bottom portion of the heat-map consists of entries not accessed by any
simulation trace; this is because the control designer indicated interest only
in fuel injection rates less than a certain amount. Thus entries corresponding
to values greater than this amount were never accessed. This also shows another
value of our tool: it allows visualizing {\em coverage} of \lut entries or the
parameter space by a given set of simulation runs.

\subsubsection{\textbf{Study of Responsiveness in a Hydrogen Fuel-Cell Vehicle}}
Next,
we use a prototype Model-In-the-Loop-Simulation (MILS) model of an
airpath control model from a hydrogen fuel-cell (FC) vehcile
powertrain. This model has more than 7000 \simulink blocks that give a
detailed description of the physics of the airpath, along with a
simplified model of the power management, and a complex controller
with several look-up maps to regulate the flow of air through the
fuel-cell stack. A key requirement of the closed-loop system is
responsiveness: how well does the system react to a driver's request
for increased torque. Internally, a torque request gets translated to
a request for increased air-flow through the stack. Thus the rise time
on the air-flow rate signal is a good proxy for system responsiveness
(STL requirement~\eqref{eq:risetime}). In the requirement, $r$ is the
rise-time, and $\lambda$ is a suitable number in $[0.5,1]$.
\begin{equation}
\label{eq:risetime}
\hspace*{-3mm}
\ffcrisetime\! \triangleq 
\alw_{[0,T]}\!\!\left(\!\step(\!\AFR_{\mathrm{ref}}\!) \Rightarrow \ev_{[0,r]}(\!\AFR \!>\! \lambda\!\cdot\! \AFR_{\mathrm{ref}})\!\right)
\hspace*{-2mm}
\end{equation}

In this case study, we choose three key controller look-up maps and
study the correlation between accessing a certain region of the
look-up map with the responsiveness of the closed-loop system. As the
model is proprietary, we suppress the values on the axes. 

\begin{figure*}[t]
\centering
\subfigure[Importance of \lut entries for the requirement related to
responsiveness of $\fcmodel$.  The results shown are w.r.t. the
$R_{\dstar}$ heuristic not incorporating the frequency of access and
metric-space interpretation of the table.]{%
    \label{fig:FC_bool}
    \begin{tikzpicture}
    \node[inner sep=0pt] (pic) {\includegraphics[height=40mm]{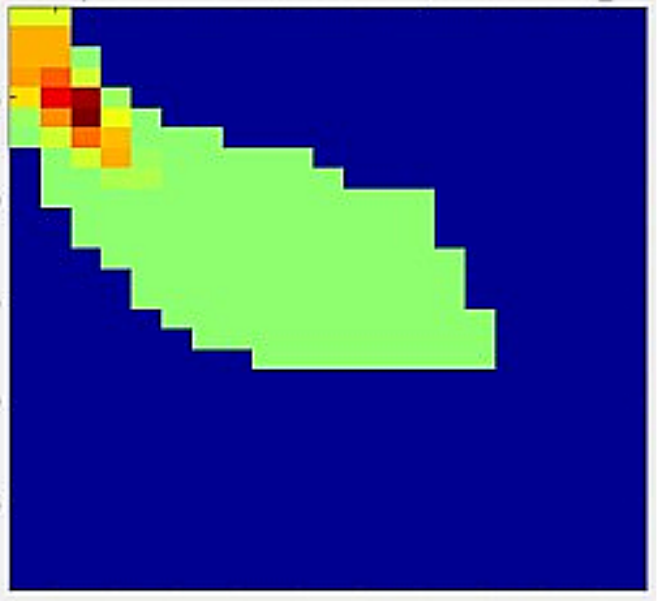}};
    \node[rectangle,font=\fontsize{8}{8}\selectfont,left of=pic,rotate=90,node distance=30mm] (blah) {Pressure Ratio}; 
    \node[rectangle,font=\fontsize{8}{8}\selectfont,below of=pic,node distance=25mm] (blah2) {Air-Flow Rate (Nlmm$^{-3}$)};
    \end{tikzpicture}
} \qquad
\subfigure[Importance of \lut entries for the requirement related to
responsiveness of $\fcmodel$.  The results shown are w.r.t. the
$R_{\dstar}$ heuristic incorporating the frequency of access and
metric-space interpretation of the table.]{%
    \label{fig:FC_freq}
    \begin{tikzpicture}
    \node[inner sep=0pt] (pic) {\includegraphics[height=40mm]{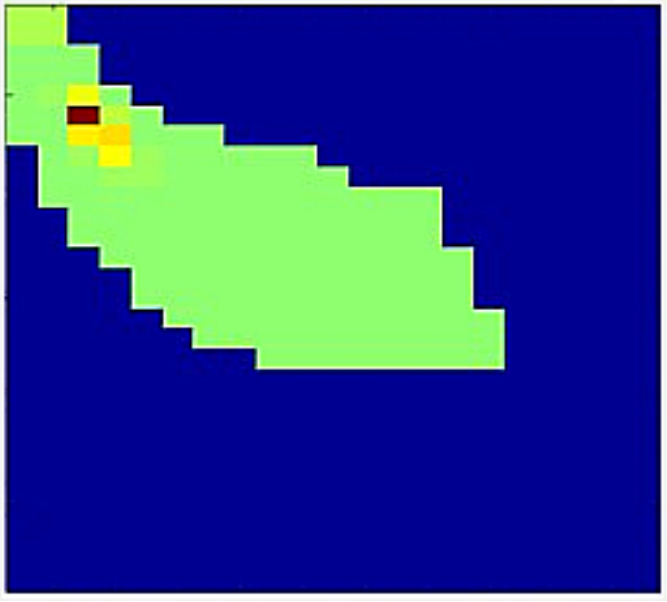}};
    \node[rectangle,font=\fontsize{8}{8}\selectfont,left of=pic,rotate=90,node distance=30mm] (blah) {Pressure Ratio}; 
    \node[rectangle,font=\fontsize{8}{8}\selectfont,below of=pic,node distance=25mm] (blah2) {Air-Flow Rate (Nlmm$^{-3}$)};
    \end{tikzpicture}
}
\caption{Heatmap representation for \lut entries for the
$\fcmodel$ model.\label{fig:fc_heatmap}}
\end{figure*}

\noindent\emph{Results.}
Figure~\ref{fig:fc_heatmap} depicts the heat-map of the LUT rankings for a
chosen \lut. All three \luts show similar heat-maps, and we only show one due to
lack of space.
The actual signals used to index the look-up maps correspond to the
pressure ratio across a compressor component and the air-flow through
the compressor. Our analysis shows that the model is not responsive
at low air-flow rates and pressure ratio values. The designers
confirmed that this analysis was of interest to them, and indicated
plans to improve the model performance for these conditions.


\section{Conclusion}
 In this paper we present a set of easy-to-compute
statistical correlation based rankings  in order to localize parameters  in
control software which may be causing undesired model behavior in controlled
cyber-physical systems. We empirically test the ranking heuristics provided by
these methods on a number of case studies that are relevant in an industrial
cyber physical systems context using our tool integrated into a
simulation-guided falsification  workflow for \simulink models. It is a
perfectly reasonable expectation of the reader that we suggest a
single ranking scheme that should be generally used or present some guidelines
to pick a ranking scheme for analysis.  Unfortunately, our experiments show that
each ranking scheme has its own merits, and any guidelines would have to rely on
deep knowledge of the model structure and dynamics.  It is arguable that the
union spectrum ranking scheme, when it gives a result should not be ignored by
the designer. For the other ranking schemes, our suggestion is to use them to
get an overall picture of the ``problem regions'' in the \lut, using graphical visualization
tools such as heat-maps.

\bibliographystyle{alpha}
\bibliography{debug}


\newpage

\appendix

\section*{Running Time for Ranking  Computation}

\smallskip\noindent\textbf{Basic Quantitative Similarity Coefficients}
Let $\myZ$ be the set of executions.
We assume that the code is instrumented such that each execution
$z$ stores a time-ordered list  of map indices that are accessed during the execution.
We denote by $|z|$ the number of function queries in
 $z$, that is, if in $z$ we get queries for
$f_M^{\interpol}(m_1), f_M^{\interpol}(m_2), \dots f_M^{\interpol}(m_p)$ (the queries need not be distinct), then $|z|$ is $p$.
We assume that the interpolation scheme uses a constant number of value to construct
an interpolation, that is we assume $\abs{\depends(m)}$ to be a constant.
Thus, the time-ordered list  of map indices that are accessed during the execution
is of size $O(\abs{z})$.
We let $ \size(\myZ)$ denote $\sum_{z\in \myZ} \abs{z}$.
Note that the total size of the time-ordered list  of map indices that are accessed during the executions in $\myZ$ is at most $ \size(\myZ)$.
The binary similarity coefficients for $m\in \dom(M)$ can be computed in time linear in the
size of this time-ordered list.
Finally, the ranking is done by sorting the indices based on the similarity coefficients.
Thus, the total time required is 
$O\left( \size(\myZ)   \ + \ \abs{M} \log\left( \abs{M}\right)\right)$.

\smallskip\noindent\textbf{Similarity Coefficients utilizing
Continuity and the Metric Space structure}
Let $\myZ$ be the set of executions.
In this case, we assume that the code is instrumented such that each execution
$z$ stores, a time-ordered list  of 
(i)~map indices that are accessed during the execution, and
(ii)~function arguments to
$f_M^{\interpol}$ that are queried for, \ie, if the execution queries
$f_M^{\interpol}(m_1), f_M^{\interpol}(m_2), \dots$, then we store
$m_1, \depends(m_1), m_2, \depends(m_2), \dots$.
From this list for $z$, we construct another list containing $\raffect(m)$ values for
every $m\in \dom(M)$, where $\raffect$ is as in Equation~\eqref{eq:raffectMetric}.
This can be done by first building another list which, for each $m_k$
in the original list (arising from a  $f_M^{\interpol}(m_k)$ query),
creates a sublist of all $\maffect(m_k, m)$ values for $m\in \dom(M)$ such that
$\maffect(m_k, m) > 0$. 
In case $r_M$ is $\infty$, this takes $O(\abs{M})$ time for each $m_k$.
In case  $r_M$ is finite, this takes $O( \abs{M}_{r_M})   $ time, where
$ \abs{M}_{r_M}$ denotes the maximum number of indices of $M$ in an
$r_M$ sized ball (in the indices space).
From this secondary list, the list of $\raffect$ values can be constructed in linear
time, and the coefficients constructed.
Thus, the total running time is 
$O\left( \size(\myZ) \cdot \abs{M}_{r_M}  \ + \ \abs{M} \log\left( \abs{M}\right)\right)$.
Thus, having a small  radius $r_M$ allows us to avoid a (possibly quadratic) blowup in the
running time as compared to the basic similarity coefficient approach.

\smallskip\noindent\textbf{Similarity Coefficients incorporating
frequency of accessing a map}
For both definitions of $\maffect$, we can proceed as follows.
Sort the arguments for  queries in each execution $z$ (\ie, sort the $m_k$ values
where $f_M^{\interpol}(m_k)$ is called in $z$), based on some ordering of the indices; and then
add up  $\maffect()$ values for the same $m_k$ values.
This step takes $O\left(\abs{z}\cdot \log\left(\abs{z}\right)\right)$ time,
and gives a list of the $\fraffect$ values for $z$.
The similarity coefficients then be calculated in linear time, and after that we need to
sort $M$ elements to get a ranking.
Thus, the total time required is 
$O\left( \sum_{z\in\myZ}\left(\abs{z}\cdot \log\left(\abs{z}\right)\right)
   \ + \ \abs{M} \log\left( \abs{M}\right)\right)$.

\smallskip\noindent\textbf{Union Spectrum method}
The sets $M_F\subseteq M$ and $M_S\subseteq M$ can be computed in time 
$O(\abs{M} + \size(\myZ))$ as a sorted list.
The set  $\ball(M_S, r_M)$ can be obtained as a sorted list
in time $O\left( \abs{M_S}\cdot  \abs{M}_{r_M}\right)$ 
where $ \abs{M}_{r_M}$  denotes the maximum number of indices of $M$ in an
$r_M$ sized ball (in the indices space).
The set difference $M_F \setminus \ball(M_S, r_M)$ can be computed time
$O\left( \abs{M_F} + \abs{\ball(M_S, r_M)}\right)$.
Putting everything together, we get that
$\sus_U$ can be computed in 
$O\left(\size(\myZ) \, +\, \abs{M}\cdot \abs{M}_{r_M}\right)$ time.
The values $s_U(m)$ can be inferred for all $m\in \dom(M)$ 
by maintaining some additional bookkeeping
in the above algorithm without increasing the running time complexity.
The values $d_U(m)$ can also be computed from
the sets  $M_F$ and $M_S$ augmented with some additional bookkeeping in
linear time.
Finally, we need to sort according the values $R_U(m)$.
The total running time works out to be
$O\left(\size(\myZ) \, +\, \abs{M}\cdot \abs{M}_{r_M}\ +\ 
\abs{M}\cdot \log(\abs{M})
\right)$.

\end{document}